\documentclass[sigconf]{acmart} 

\usepackage{booktabs} 
\usepackage{balance} 

\begin{document}

\copyrightyear{2019} 
\acmYear{2019} 
\setcopyright{acmlicensed}
\acmConference[LAK19]{The 9th International Learning Analytics & Knowledge Conference}{March 4--8, 2019}{Tempe, AZ, USA}
\acmBooktitle{The 9th International Learning Analytics \& Knowledge Conference (LAK19), March 4--8, 2019, Tempe, AZ, USA}
\acmPrice{15.00}
\acmDOI{10.1145/3303772.3303798}
\acmISBN{978-1-4503-6256-6/19/03}

\title{Towards Value-Sensitive Learning Analytics Design}

\author{Bodong Chen}
\orcid{0000-0003-4616-4353}
\affiliation{%
  \institution{University of Minnesota}
  \streetaddress{159 Pillsbury Drive SE}
  \city{Minneapolis}
  \state{MN, USA}
  \postcode{55455}
}
\email{chenbd@umn.edu}

\author{Haiyi Zhu}
\affiliation{%
  \institution{University of Minnesota}
  \streetaddress{200 Union St SE}
  \city{Minneapolis}
  \state{MN, USA}
  \postcode{55455}
}
\email{zhux0449@umn.edu}

\renewcommand{\shortauthors}{Chen \& Zhu}

\begin{abstract}
To support ethical considerations and system integrity in learning analytics, this paper introduces two cases of applying the Value Sensitive Design methodology to learning analytics design. The first study applied two methods of Value Sensitive Design, namely stakeholder analysis and value analysis, to a conceptual investigation of an existing learning analytics tool. This investigation uncovered a number of values and value tensions, leading to design trade-offs to be considered in future tool refinements. The second study holistically applied Value Sensitive Design to the design of a recommendation system for the Wikipedia WikiProjects. To proactively consider values among stakeholders, we derived a multi-stage design process that included literature analysis, empirical investigations, prototype development, community engagement, iterative testing and refinement, and continuous evaluation. By reporting on these two cases, this paper responds to a need of practical means to support ethical considerations and human values in learning analytics systems. These two cases demonstrate that Value Sensitive Design could be a viable approach for balancing a wide range of human values, which tend to encompass and surpass ethical issues, in learning analytics design.
\end{abstract}

%
%




\maketitle

\section{Introduction}

Over the past few years, considerations of the ethical implications of learning analytics have become a central topic in the field \cite{Prinsloo2017-tn}. The increasing emphasis on ethical considerations in learning analytics is situated within growing public concerns with algorithmic opacity and bias, as well as awakening advocacy of algorithmic transparency and accountability \cite{ACM_US_Public_Policy_Council2017-sy}. In the field of learning analytics, a rich body of theoretical, practical, and policy work has emerged to promote the ethical collection and use of learning data. In 2016, the \emph{Journal of Learning Analytics} published a special section featuring work on ethics and privacy issues \cite{Ferguson2016-qw}. Ethical frameworks, codes of ethical practice, and ethical consideration checklists have been developed \cite{Sclater2016-mn, Prinsloo2016-ic, Drachsler2016-zb}. New policy documents are also put forward by higher education institutions to enhance transparency with regard to data collection and data usage.\footnote{See the SHEILA project for examples: http://sheilaproject.eu/la-policies/}

In contrast with the private sector, education as a social function also has a moral responsibility to promote student success. Therefore, ethical considerations in learning analytics need to go beyond issues that are foregrounding the current public discourse surrounding surveillance and privacy \cite{Ferguson2016-qw}. Educational institutions have ``an obligation to act''---to effectively allocate resources to facilitate effective teaching and learning through ethical and timely interventions \cite{Prinsloo2017-jv}. Ethical practice in learning analytics entails considerations of not only typical ethical issues but also factors and values pertinent to the larger education system. Recently, Buckingham Shum \cite{Buckingham_Shum2017-un} proposes a \textit{Learning Analytics System Integrity}  framework that goes beyond algorithmic transparency and accountability to consider multiple pillars of a learning analytics system including learning theory, algorithm, human--computer interaction, pedagogy, and human-centered design. While learning analytics as a field is rightfully giving more attention to ethical challenges with data analytics, to foster ethical practice we also need effective methods to consider ethical challenges in a manner that would engage perspectives from different stakeholders while allowing education---as a complex system---to fulfill its societal function.

To this end, this paper proposes to apply \emph{Value Sensitive Design}---a theory and methodology initially developed in the field of Human--Computer Interaction (HCI)---to support the design of value-sensitive learning analytics systems. We posit that Value Sensitive Design can (1) guide researchers and practitioners in the process of evaluating and refining existing learning analytics applications, and (2) provide systematic guidance on developing new learning analytics applications to more holistically address values pertinent to stakeholders, educators, and the society.

The remainder of this paper is structured as follows. We first briefly review work on ethics frameworks, algorithmic transparency, and accountability in the field of learning analytics. Then we turn to the Value Sensitive Design literature, with a focus on its application to the development of information systems. After outlining our research goals, we introduce two studies in Sections~\ref{sec:study1} and \ref{sec:study2}. We conclude this paper by discussing the importance of considering values and value tensions in learning analytics design and development. 

\section{Background}

\subsection{Learning Analytics Systems: Ethics and Values}

Learning analytics is defined as ``the measurement, collection, analysis and reporting of data about learners and their contexts, for purposes of understanding and optimising learning and the environments in which it occurs'' \cite[p.~34]{Long2011-ky}. By harnessing increasingly abundant educational data and data mining algorithms, learning analytics aspires to build applications to personalize learning based on individual progress, predict student success for timely intervention, provide real-time feedback on academic writing, and so forth \cite{Siemens2013-az}. As learning analytics applications are increasingly integrated in the existing academic technology infrastructure (e.g. student information systems, learning management systems, student support systems), more educational institutions are becoming equipped with strong, connected analytics capacity to better understand students and hopefully to also better support student success. Empirical evidence accumulated so far has indeed demonstrated promise of learning analytics in supporting student learning \cite{Huberth2015-fd}.

While learning analytics are garnering international interests across different levels of education, ethical and privacy concerns are increasingly mentioned in both public and scholarly discourse. Slade and Prinsloo \cite{Slade2013-xt} listed a number of ethical challenges in learning analytics, including the location and interpretation of data; informed consent, privacy, and de-identification of data; and the classification and management of data. Scholars have organized a series of workshops at the Learning Analytics and Knowledge (LAK) conference to collectively tackle ethical challenges \cite[e.g. ][]{Drachsler2016-wf}, leading to the development of ethical principles, frameworks, and checklists to guide ethical research and practice \cite{Willis2016-kh, Drachsler2016-zb, Ferguson2016-qw, Rodriguez-Triana2016-km}. Empirical work has also been done, for example, to examine student perceptions of privacy issues with regard to learning analytics applications \cite{Ifenthaler2016-qd}. Design work following the \textit{Privacy by Design} principles has shown promise in identifying solutions to addressing privacy concerns \cite{Verhagen2016-sf}. While institution-level code of ethics are being created, efforts are also made to help individual practitioners apply abstract principles and code of ethics in real-world scenarios \cite{Lang2018-sa}.

Undergirding these discussions of ethics and privacy are \emph{values} in learning analytics systems. Generally speaking, \textit{value} represents either ``something important, worthwhile, or useful'' or ``one's judgment of what is important in life'' (Oxford English Dictionaries, 2018). For instance, privacy is a fundamental human right, which holds not only value for a person but also social and public values for democratic societies \cite{Regan1995-ht}. So safeguarding privacy has naturally become an important ethical consideration. There are other values that are also important in learning analytics but often go unstated. For instance, as educators we value student success, learner agency, and equitable access to education \cite{Ferguson2016-qw}, but these values, despite their educational significance, are often neglected in the discussion of data ethics. In other words, various values are not equally recognized in current discourse, leaving some values and tensions among values neglected or under-explored.  To promote ethical practice in learning analytics, we need means to interrogate the values that are often in tension with each other. The learning analytics \textit{accountability analysis} put forward by Buckingham Shum \cite{Buckingham_Shum2017-un} considers system integrity from multiple angles and holds promise for eliciting values beyond current thinking on ethical issues. Take automated feedback on academic writing for example. Values supported by an academic writing analytics tool include the \textit{accurate} detection of rhetorical moves based on linguistic features and \textit{usable} feedback interface that can facilitate student sense-making. However, the accurate detection of rhetorical moves may demand more learner data and is hereby in tension with the value of privacy foregrounded in ethical considerations. To better support learning analytics design, the field needs strategies for weighing competing values and deriving design trade-offs to mitigate value tensions. 

\subsection{Value Sensitive Design}

Stemming from an orientation towards human values, \emph{Value Sensitive Design} ``represents a pioneering endeavor to proactively consider human values throughout the process of technology design'' \cite[p.~1]{Davis2013-eo}. In Value Sensitive Design, value is operationally defined as ``what is important to people in their lives, with a focus on ethics and morality'' \cite[p.~68]{Friedman2017-ti}. Value Sensitive Design offers a systemic approach with specific strategies and methods to help researchers and designers explicitly incorporate the consideration of human values into design \cite{Davis2013-eo, Friedman2017-ti}. It recognizes the complexity of social life and attempts to illuminate \emph{value tensions} among \emph{stakeholders}. 

Value Sensitive Design is a theory, a methodology, and an established set of methods \cite{Davis2013-eo}. As a methodology, Value Sensitive Design involves three types of investigations: conceptual, empirical, and technical \cite{Davis2013-eo}. Together they could form an iterative process of identifying and interrogating values in technology design. According to \cite{Friedman2013-hx,Davis2013-eo}, conceptual investigations are primarily concerned with identifying direct and indirect stakeholders and eliciting values held by different stakeholders. A conceptual investigation can draw from philosophical and theoretical texts in order to conceptualize a value at hand (e.g. trust, autonomy). Empirical investigations venture further to collect empirical data from stakeholders about their perceptions of a value and value tensions (e.g. privacy vs. utility) in a specific context. Both qualitative and quantitative methods can be used in empirical investigations. Finally, technical investigations focus on the designed technology itself and seek to either proactively design new technological features to support values or examine how a current design would support or undermine human values. These three types of investigations can be integrated in an iterative design process to holistically address human values.

Fourteen different methods can be drawn to support Value Sensitive Design \cite{Friedman2017-ti}. For example, \textit{direct and indirect stakeholder analysis} can lead to the identification of indirect stakeholders of transit information tools \cite{Watkins2013-yn}; fictional \textit{value scenarios} are useful for revealing values and tensions when developing mobile apps \cite{Czeskis2010-pi}; \textit{value dams and value flows} could be a useful analytic method for resolving value tensions among design choices---by removing a design choice based on strong objection from a small percentage of stakeholders (the value dams) or accepting a design option based on favorable reaction from a certain portion of stakeholders (the value flow) \cite{Czeskis2010-pi}. These methods suit varied purposes of design and a project adopting Value Sensitive Design can draw from multiple methods. 

\subsection{Overview of the Paper}

To examine how \textit{Value Sensitive Design}---the theory, methodology, and methods---could be applied to learning analytics design, this paper reports on two studies.

The first study is a conceptual investigation of a learning analytics application that has already been developed. Applying two Value Sensitive Design methods---stakeholder analysis and value analysis \cite{Friedman2017-ti}---this study aimed to (1) identify key stakeholders of the developed application, (2) elicit values and value tensions, and (3) derive design trade-offs to be considered in future design iterations. 

Moving beyond conceptually investigating an existing system, the second study introduces a case of applying Value Sensitive Design to proactively consider human values throughout a design process. The introduced case is about developing a value-sensitive recommendation system for the Wikipedia WikiProjects. Applying ideas from Value Sensitive Design, we and colleagues (1) engaged stakeholders in the early stages of the algorithm design and used stakeholders' insights to guide the algorithm development, (2) iteratively improved, adapted, and refined the algorithmic system based on the stakeholders' feedback, and (3) evaluated the recommendation system not only based on accuracy but also stakeholders' acceptance and their perceived impacts of the system.

\section{Study I: A Conceptual Investigation of a Discussion Visualization Tool}
\label{sec:study1}

\subsection{Study Context}

The first study was a conceptual investigation of a learning analytics tool developed to visualize online discussions in a learning platform named Yellowdig.\footnote{See https://yellowdig.com/} This study and the creation of Yellowdig were contextualized within higher education's interests in fostering social learning and peer interactions among students. Branded as a social media platform designed for education use, Yellowdig resembles many of social network sites such as Facebook and Twitter. For example, it features a news feed similar to that of Facebook; learners can contribute text or multimedia \textit{posts} to the news feed, where other learners can ``like'' or comment on each other's posts
. By providing social-media features students are already familiar with, this tool aims to facilitate social learning in online and blended classes without intruding into students' personal social networks. Since its inception in 2014, Yellowdig has been adopted by a number of higher education institutions.


As one of the early adopters of Yellowdig, Northwestern University developed a \textit{Yellowdig Visualization Tool} to visualize students' peer interactions and conceptual engagement on Yellowdig.\footnote{See https://www.teachx.northwestern.edu/projects2016/gruber/} 
The design of this tool was similar to many existing social learning analytics tools such as SNAPP \cite{Dawson2010-ws}, Threadz,\footnote{https://threadz.ewu.edu/} and CavanNet \cite{Chen2018-mc}. The tool was chosen for this study because it represented a more recent design effort, has a fairly polished design, and has been informed by participatory design workshops attended by students at the university. In other words, this tool could well represent the state-of-the-art of learning analytics applications designed for this particular context. 

\begin{figure}
\includegraphics[width=0.5\textwidth]{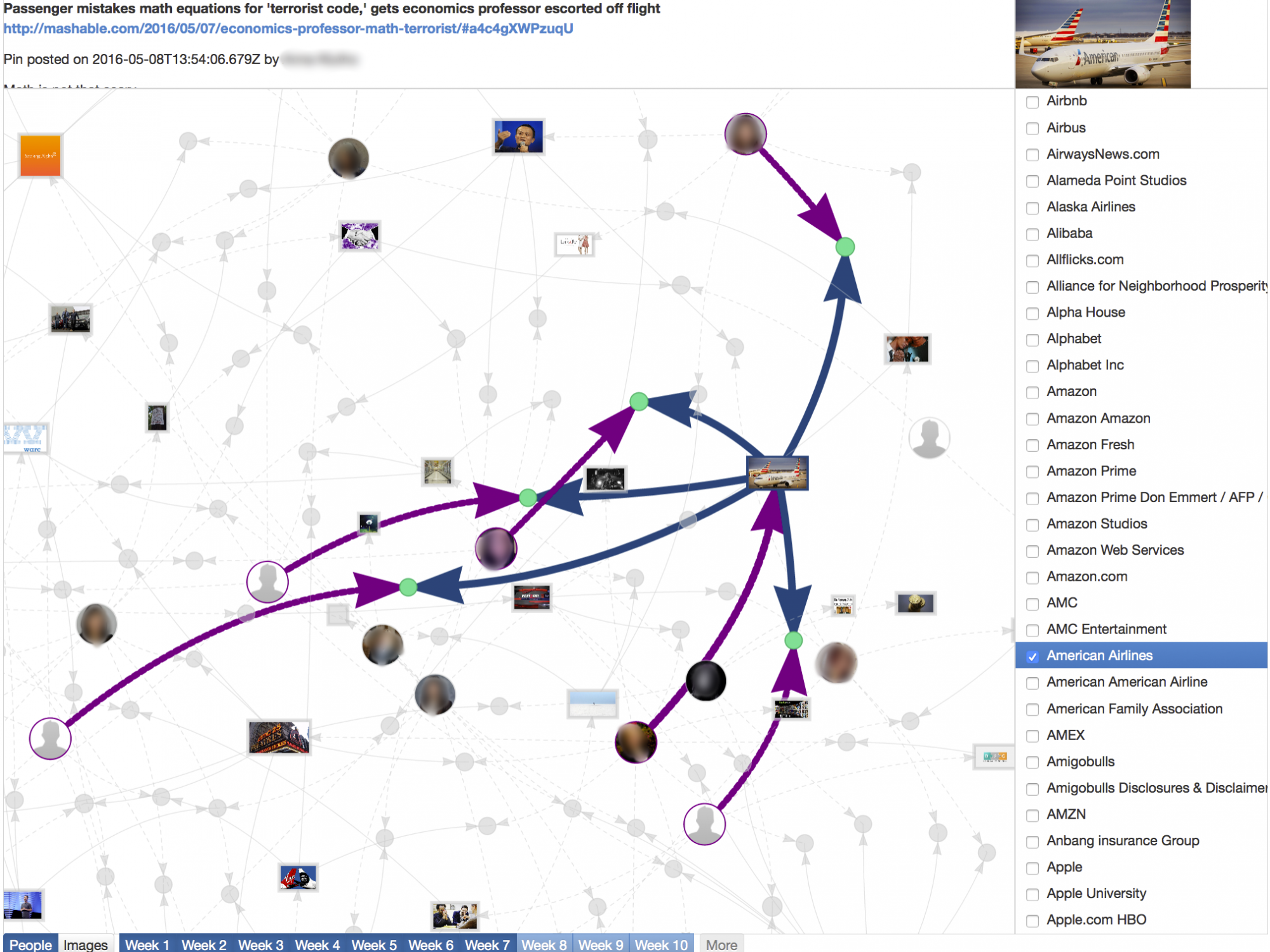}
\caption{Yellowdig Visualization Tool. (\emph{Note}: Used with permission of the Yellowdig Visualization development team at Northwestern University.)}
\label{fig:vis}
\end{figure}

As described by its project website, the development of the Yellowdig Visualization Tool was motivated by an interest in helping students and faculty learn more about their online discussions by accessing Yellowdig data. Prior research on educational use of Web 2.0 tools and online discussion environments has demonstrated the need to support learner participation and engagement from pedagogical, technological, and sociocultural angles \cite{Greenhow2017-qq, Czerkawski2016-au}. The Yellowdig Visualization Tool was one example of supporting student use of Yellowdig through designing a new learning analytics application, which is expected to be coupled with pedagogical interventions. During the design process, the design team came up with an initial prototype and engaged students in participatory design to elicit their ideas about visualizations that could be useful for their learning. The version of the tool analyzed in this study contained the following key features (see Figure~\ref{fig:vis}):

\begin{itemize}
\item \textit{Multi-mode, multi-plex network visualization.} The visualization tool represents one class's Yellowdig discussion as a multi-mode, multi-plex network that contains multiple types of nodes (students, posts, and comments) and multiple types of relations (posting, commenting). In this network, each Yellowdig post (also called ``pin'') is represented by a blue square filled either by solid blue color (if the post contains pure text) or a thumbnail (if the post contains a rich media object); comments on posts are shown as light green circles; users are represented by grey user icons. Following a typical network visualization heuristic, the size of a post nodes is correlated with the number of connections it has. The whole network is plotted using a force-directed layout, which keeps well connected nodes in the center and pushes less connected nodes to the edge. 
\item \textit{Node highlighting.} When a node is chosen by the user, the visualization would highlight the node's ``2.0 level ego network,'' which comprises the chosen node (i.e. the ego) and all other nodes that can be reached within two steps. The content of the chosen node is also displayed in the top panel of the tool (see  Figure~\ref{fig:vis}). 
\item \textit{Visualization of temporal growth.} A movie can be played to inspect both daily and weekly growth of the discussion network. The placement of all nodes and edges are fixed in advance. The temporal visualization adds nodes and edges into the network in a step-wise manner.
\item \textit{Filtering mechanisms for instructor use}. The instructor view of this visualization tool provides additional mechanisms for filtering the network by student names, post tags, and named entities (such as Uber) recognized by text mining algorithms. These filters are displayed on the right panel of the visualization for the instructor to explore the discussion network.
\end{itemize}

In comparison with earlier social learning analytics applications, the Yellowdig Visualization Tool is novel in several ways. First of all, in comparison with one-mode networks adopted in most social network analytics, the Yellowdig Visualization Tool turns discussion data into a multi-mode, multiplex, temporal network that provides a more holistic picture of student discussions. Its node highlighting feature allows students and the instructor to inspect local network structures, and its filtering mechanisms support the instructor to quickly identify discussion trends. To a great extent, the visualization tool provides a unique view of the Yellowdig discussion, which can facilitate sense-making activities by its users and serve assessment needs of the instructor.

\subsection{Methods}

To study values in such a nascent learning analytics tool, we conducted a conceptual investigation \cite{Davis2013-eo} to identify its stakeholders and elicit values and value tensions among identified stakeholders. 
This conceptual investigation had two components: 

\begin{enumerate}
\item \textit{Stakeholder analysis}. The first component was to identify the direct and indirect stakeholders of this tool. According to \cite{Friedman2017-ti}, this analysis includes individuals who are in direct contact with this tool (\textit{direct stakeholders}) and those who may not interact with the tool directly but are likely to be impacted by tool usage (\textit{indirect stakeholders}). 
\item \textit{Value analysis}. The second component was to identify and define the benefits, harms, and values implicated by this learning analytics tool and to illuminate value tensions embedded in the tool and its larger education systems. Values at stake in this analysis include generic human values that represent ``what is important to people in their lives" \cite[][p.~68]{Friedman2017-ti}, as well as values endorsed by the educational literature.
\end{enumerate}

This conceptual investigation was grounded in our knowledge of higher-education classes in similar settings and aided by technical inspection of this Yellowdig Visualization Tool. It is expected that through this study we could come up with suggestions for considering identified values and design trade-offs in future iterations of this particular tool as well as other similar applications.

\subsection{Findings}

\subsubsection{Stakeholders}

In this context, the \textit{direct stakeholders} included students and instructors accessing the Yellowdig Visualization Tool. It was expected that the tool would assist instructors in analyzing the data generated on Yellowdig to inform their instructional decisions. The tool design also aimed to enhance students' mastery of the course content and provide them with another way of engaging with the course material and with each other (see project website). However, the adoption of such a learning analytics tool is often decided by the instructor and its acceptance among students may vary. Therefore, students may be further divided into those who choose to use the tool and those who reject the tool for various reasons. 

Because Yellowdig is a private social media platform, the visualization tool is based on data generated by the class and is not accessible beyond the class. The \textit{indirect stakeholders} of this tool may include future cohorts of the class whose discussion performance may be compared (explicitly or implicitly) with the present cohort. Academic technology staff members are also indirect stakeholders as they are charged with supporting technology use and may be asked to provide additional technical support for this tool. Student support services, such as academic advisers and academic analytics teams, are also indirect stakeholders. For example, an academic adviser may be contacted by the instructor when she finds from the tool that a student has been inactive for a few weeks.

\subsubsection{Values and tensions}

The Yellowdig Visualization Tool aimed at providing another way of engaging with the class discussion, enhancing student learning and social interaction, and assisting the instructor to grasp discussion content in order to make informed instructional decisions. However, through the conceptual investigation we found these values supported by the tool can be in tension with other values. 

For such a learning analytics tool that can be used for assessment purposes, student \textbf{autonomy} is an important value that is in tension with the tool's \textbf{utility} for assessment. The network visualization and node highlighting features of the tool make it easy to identify posts and comments made by each individual student. Even though such information is also accessible via the original Yellowdig discussion interface, the visualization tool makes it more readily accessible to all members of the class. These design features are meant to support the values of \textbf{student success}, \textbf{accountability}, \textbf{discussion engagement}, and \textbf{usability}, but they can also press students to participate more actively and hereby hinder their autonomy as discussion participants. 

In a same vein, because of the aggregation of student participation data, student \textbf{privacy} as another important value is also at tension with these intended benefits. Similarly, even though the visualization tool does not collect new data from students, it reveals a bird-eye view of each student's participation patterns and could appear intruding to some students who are less comfortable with such aggregated reporting.

\textbf{Social well-being}, a sense of belonging and social inclusion, is another important value to be considered in this context. Students in the class---using the tool or not---can reasonably expect to be free from embarrassment caused by less active participation or less central status in the network. The force-directed layout adopted to generate the network visualization pushes less active participants to the edge and make them more identifiable by members of the class. This design is intuitively understandable, is broadly applied in network visualization, and supports \textbf{pedagogical decisions} by revealing to the instructor students ``at risk'' of dropping out of Yellowdig discussion. Nonetheless, students on the edge may feel embarrassed among peers should their identities be revealed. This layout can potentially contribute to the ``rich-club'' phenomenon in online discussions \cite{Vaquero2013-ld}, potentially channeling more attention to students in the center and leading to the exclusion of less active students in future interactions. 

There are also values that are implicated by algorithmic decisions embedded in the visualization tool, including \textbf{freedom from bias} and \textbf{self-image}. For instance, the force-directed layout algorithm inherently boost the status of students who are more connected in terms of connection volumes and do not consider the quality of their posts. Displaying the thumbnails of discussion posts renders posts with multimedia objects more attractive in comparison with posts with pure text, and can hereby direct varied attention to posts with and without multimedia objects. Also notable is the choice of scaling posts based on the number of comments, which directs more attention to posts that have already attracted more attention. These differentiated treatments of discussion posts embedded in the algorithms are linked to students' \textbf{self-image}. When a student's node gets highlighted in the network visualization, the highlighted ego network essentially represents what the student has produced and how his/her participation is linked with peers in the community. Being able to see all connections of a student at once can benefit the \textbf{sense of community} especially when a student see herself well connected. However, it may also pose questions to a student's self-image if she is less connected with peers in the community. 

The visualization tool was designed to provide another way to make sense of and engage in online discussions on Yellowdig. Values facilitated by the tool include \textbf{utility}, \textbf{usability}, and \textbf{ease of information seeking} in Yellowdig. In particular, the tool provides views of both the overall discussion structure and micro, local interaction patterns that are not revealed by the Yellowdig interface. The instructor view provides rich filtering mechanisms that are useful for filtering discussion posts by students, tags, and/or named entities. Being able to look at entity names, such as brands in a business class, allows the instructor to quickly identify popular topics in student discussions and make pedagogical decisions on, for example, whether to address a specific topic in class. However, the fact that these advanced features are reserved for the instructor is at tension with the values of \textbf{epistemic agency} and \textbf{fairness}. One can argue that the advanced features can equally benefit students, who also need to complete information seeking tasks to effectively participate in Yellowdig discussions. Nevertheless, one may also argue that providing these additional features places unnecessary cognitive overload and \textbf{burden} on students. 

\subsection{Discussion and Implications}

With aid from technical analysis of the Yellowdig Visualization Tool, the conceptual investigation has revealed a number of values and value tensions that have implications for future refinements of the tool. One premise of discussing the implications is that the accommodation of these values should not be solely achieved through technology design, but also the design of pedagogical interventions \cite{Wise2017-yn}.  Nonetheless, based on findings from the value analysis, we suggest the following possible design trade-offs that could be considered in future technology design:

\begin{itemize}
    \item Considering the tension between \textbf{utility} (e.g. revealing a student's participation) and the values of students' \textbf{autonomy}, \textbf{privacy}, \textbf{social well-being}, and \textbf{self-image}, one design idea is to give each student the option of not revealing his/her name to their peers in the network visualization. In this way, students can \textit{choose} to become less identifiable in the tool. Another possible idea is to make student nodes non-clickable so that the node highlighting feature is only limited to post and comment nodes. Both design ideas will sacrifice utility of the visualization tool to support other values such as privacy and social well-being. 
    \item To support the values of \textbf{freedom from bias} and \textbf{self-image}, future implementation of the network visualization could consider adjusting parameters of the force-directed layout or consider other graph drawing algorithms (such as the circular, arc diagram, hive plot, and hierarchical graph layouts). Here, design trade-offs are likely to involve interpretability of a network visualization and its complications concerning student status in the given sociocultural context. Empirical investigations of student perceptions of different layouts could be necessary.
    \item Considering the values of \textbf{fairness}, \textbf{epistemic agency}, and \textbf{ease of information seeking}, the filtering mechanisms provided to the instructor could be also considered for student access. Given the tension between \textbf{privacy} and \textbf{ease of information seeking}, filtering by student names could still be reserved to the instructor, while the other two filtering mechanisms (i.e. by tags or named entities) can be made available for students.
\end{itemize}

The design ideas mentioned above are presented here to illustrate the promise of Value Sensitive Design for the particular context, instead of disproving the current visualization tool that is quite well-designed. These design ideas are tentative and by no means comprehensive. Further empirical and technical investigations are necessary in order to advance these design ideas. 

\section{Study II: A Holistic Application of Value Sensitive Design to an Algorithmic System of WikiProjects}
\label{sec:study2}

\subsection{Study Context}

While the first study focuses on social learning in formal classes, the second study was situated in the informal context of Wikipedia editing. Specifically, this study was concerned with designing recommendation algorithms for group formation in Wikipedia's WikiProjects. A WikiProject organizes ``a group of participants... to achieve specific editing goals, or to achieve goals relating to a specific field of knowledge.''\footnote{See https://en.wikipedia.org/wiki/WikiProject} Editing a Wikipedia article is itself a learning and knowledge-creation process. Learning scientists have recognized mass collaboration, on Wikipedia for instance, as an emerging paradigm of education \cite{Cress2016-kv}. Computer-supported collaborative learning (CSCL) researchers have studied the social process of building collective knowledge on Wikipedia, leading to findings about the intricate relations between contributors and Wikipedia articles \cite{Halatchliyski2014-rc}. Indeed, coordinated editing in a WikiProject involves a great deal of sense-making, negotiation, synthesis, and knowledge transformation. For this reason, since its inception Wiki as a Web 2.0 technology has been widely adopted in formal classrooms to support sophisticated collaborative learning experiences \cite{Greenhow2017-qq}. Therefore, facilitating knowledge processes on the Wiki technology---and on Wikipedia in particular---is of interest to learning analytics because Wikipedia editing entails learning, and intelligent supports built for Wikipedia are transferable to Wiki-based learning scenarios. 

Prior work has illuminated that Wikipedia, like other social media environments, now heavily depends on data-driven algorithmic systems to support its massive-scale collaboration and governance \cite{Geiger2017-fl}. 
On Wikipedia, algorithmic systems have been used to support a variety of critical tasks such as counter-vandalism, task routing, and the Wikipedia education program.
Many of these existing algorithmic systems are driven by ``Big Data'' and supervised machine learning algorithms. Take predicting vandalism for example. The first step of the task is to define a prediction target, i.e. a malicious edit. The second step in the process is to use historical data, often in large volumes, to train and validate machine learning models. Finally, the validated models are applied to new edits in order to generate predictive scores of being malicious edits. Such algorithmic systems play important roles in the Wikipedia ecosystem.

However, the data-driven approach can lead to unintended biases in algorithms and potentially negative impacts on the editor community. It is recognized that the data-driven approach relies largely on historical human judgments, which are subject to historical stereotypes, discrimination, and prejudices. Using historical data to inform the future runs the risk of reinforcing and repeating historical mistakes and thus fails to capture and incorporate human insights on how the system can be improved for the future. Recent ethnographic work of Wikipedia has revealed critical issues with current algorithmic systems and calls for the study of sociocultural processes, such as newcomer socialization, that are strongly impacted by the automated systems \cite{Geiger2017-fl}.

Researchers have called for the development of systematic methods to detect and mitigate biases in existing algorithmic systems \cite{Sandvig2014-jt}. In contrast with a reactive approach that attempts to address bias when it gets exposed, we argue for an approach that would proactively consider human values \textit{throughout} the process of algorithm design in a principled and comprehensive manner. One possible way of achieving this goal is to apply principles of Value Sensitive Design \cite{Friedman2017-ti} to the creation of each algorithmic system. This holistic approach, explicated in great detail elsewhere \cite{Zhu2018Value}, engages stakeholders in the early stages of algorithm development and incorporates stakeholders' tacit knowledge and insights into the creation of an algorithm system. By doing this, it is hoped we could mitigate biases embedded in design choices and hence avoid undermining important stakeholder values. 

In this section, we briefly introduce a case of applying Value Sensitive Design to designing a recommendation engine for group formation in Wikipedia's WikiProjects. Similar recommendation tasks are also explored in the MOOC (massive open online course) and collaborative problem-solving contexts \cite{Rose2015-ub, Dowell2018-ig}. In contrast with Study I, this study was focused on the recommendation algorithm, a key component of a learning analytics system \cite{Buckingham_Shum2017-un}. By doing so, we hope to illustrate the possibility of attending values when building an algorithmic system's sub-component that is crucial but not observable from the user interface. Below, we explain the process of applying Value Sensitive Design to the development of the recommendation engine, and discuss the implications for the design of learning analytics systems.

\begin{figure*}[ht]
  \centering
  \includegraphics[width=\textwidth]{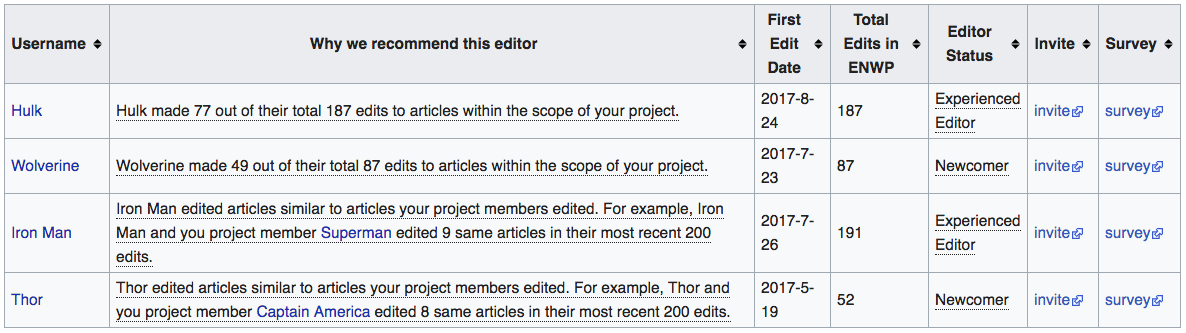}
  \caption{The interface of the prototype newcomer recommendations for WikiProject organizers.}
  \label{fig:study2_screenshot}
\end{figure*}

\subsection{Methods}
\subsubsection{Study Context}

Retaining and socializing newcomers is a crucial challenge for many volunteer-based knowledge creation community such as Wikipedia. The number of active contributors in Wikipedia has been declining steadily since 2007, due at least in part to a sharp decline in the retention of new volunteer editors \cite{halfaker2013rise}. WikiProjects, which are groups of contributors who work together to improve Wikipedia, serve as important socialization hubs within the Wikipedia community. Prior work \cite{forte2012coordination, zhu2012organizing} suggests that WikiProjects provide three valuable support mechanisms for new members: (1) enabling them to locate suitable assistance and expert collaborators; (2) guiding and structuring their participation by organizing tasks, to-do lists, and initiatives such as ``Collaborations of the Week;'' and (3) offering new editors ``protection'' for their work, by shielding it from unwarranted reverts and ``edit wars.''

However, recommending new editors to WikiProjects is not a trivial task. In the English version of Wikipedia alone there are currently about 2,000 WikiProjects, with on average 38,628 new users/editors registered to participate each month. The goal of the case study was to create algorithmic tools to recommend new Wikipedia editors to existing WikiProjects.

\subsubsection{Value Sensitive Design Process}

Drawing on the principles of Value Sensitive Design \cite{Friedman2017-ti},
we derived the following five-stage process to proactively accommodate values in the WikiProjects recommendation system. 

\begin{enumerate}
    \item Review literature and conduct empirical studies to understand relevant stakeholders' values, motivations and goals, and identify potential trade-offs.
    \item Based on the results of the first step, identify algorithmic approaches and create prototype implementations.
    \item Engage and work closely with the stakeholders to identify viable means to deploy the developed algorithms, recruit participants, and gather user feedback.
    \item Deploy the algorithms, gather stakeholder feedback, and refine and iterate as necessary.
    \item Evaluate stakeholders' acceptance, algorithm accuracy, and impacts of the algorithms on relevant stakeholders.
\end{enumerate}

In contrast with Study I that was focused on a conceptual investigation, this five-stage process incorporates all three types of Value Sensitive Design methods---conceptual, empirical, and technical---to synergistically address human values. In the proposed process, each component itself should be familiar to HCI scholars. 
However, the derived process represents an attempt to put multiple Values Sensitive Design methods into action to collectively consider values. 

\subsection{Results}

Following the described five-stage process, we have developed the algorithmic system to recommend new Wikipedia contributors to existing Wikipedia projects.

\subsubsection{Step 1: Review literature and conduct empirical studies to understand stakeholders' values and goals of importance to them, and potential trade-offs.} 

To achieve the design goal, we reviewed prior research and conducted a survey study with 59 members from 17 WikiProjects in order to answer the following questions: (1) Which factors would motivate new editors to participate in a WikiProject? (2) Which factors would motivate WikiProjects organizers to recruit new members? (3) What collective outcomes are important for WikiProjects and for Wikipedia in general?

Findings from the survey study included:
\begin{itemize}
    \item Both newcomers and WikiProject organizers valued the \textbf{interest match} (i.e. the match between the editor's interests and the project topic) and \textbf{personal connections} (i.e. the relationships between newcomers and current members of the community).
    \item WikiProject organizers valued \textbf{productivity} and wanted to recruit newcomers who could produce high-quality work;
    \item Although WikiProjects organizers valued editors' prior experience and were more interested in recruiting newcomers who already had some editing experience in Wikipedia, the Wikipeida community as a whole was more concerned with \textbf{socializing ``brand-new'' editors}.
    \item WikiProject organizers hoped to remain \textbf{control} of the recruitment process. Specifically, they objected to the idea of complete automation, i.e. to automatically identify and invite newcomers deemed relevant by a recommendation system. They wanted to have the final say on whom to be invited to their projects.
\end{itemize}  

\subsubsection{Step 2: Identify algorithmic approaches and develop system prototypes.}

Drawing on the results of the first step, the team developed recommendation algorithms that would meet the following criteria that were informed by identified stakeholders' values.

First, the recommendation algorithms needed to satisfy the goals of new editors and WikiProjects by considering the match between the editor's interests and the project topic. To this end, we created four different recommendation algorithms---two interest-based algorithms and two relationship-based algorithms.

\begin{itemize}
    \item Interest-based algorithms rank candidate editors based on how closely their editing history matches the topic of a WikiProject. Two types of interest-based algorithms were considered. A \emph{rule-based algorithm} ranks the match of an editor to a WikiProject by counting the number of edits by that editor to articles within the scope of the project. A \emph{category-based algorithm} ranks editors by computing a similarity score between an editor's edit history and the topic of a WikiProject. 
    \item Relationship-based algorithms rank candidate editors based on relationships with current members of a WikiProject. Two types of relationship-based algorithms were developed. A \emph{bonds-based algorithm} ranks editors by the strength of ``social connections'' the editor has to current members of a WikiProject. A \emph{co-edit-based algorithm} is a version of collaborative filtering and is inspired by the design of SuggestBot \cite{cosley2007suggestbot}. Candidate editors are ranked by the similarity of their edit histories to the edit histories of current members of a WikiProject.
\end{itemize}

Second, the recommendation algorithms needed to balance the interests of WikiProjects and the collective goal of Wikipedia community by targeting both experienced editors and ``brand-new'' editors. Specifically, the candidate editors included both the editors who were new to Wikipedia and editors who were moderately experienced (but not highly experienced). This design instantiated the tension we identified previously. Moreover, we also ranked the two types of newcomers separately, so that experienced ones did not overshadow the inexperienced ones.

Finally, the recommendation algorithms should satisfy the goals of WikiProjects and their organizers by excluding editors likely to produce low-quality work.

With these developed algorithms, we further designed a user interface to present top recommendations from each of the four algorithms to WikiProject organizers (see Figure~\ref{fig:study2_screenshot}). The decision of presenting multiple results was based on a finding from Step 1 that some project organizers strongly objected to the idea of complete automation and wished to ``remain in the loop'' of inviting newcomers. With the prototype interface, WikiProject organizers could review the recommendations and decide who they would like to invite to their projects \cite{Zhu2018Value}. As such, their desire to remain in the loop was accommodated. 

\subsubsection{Step 3: Engage and work closely with the community.}

Online communities like Wikipedia constitute a rich laboratory for research. Digital traces of collaboration, social interaction, and production are made visible, offering rich opportunities for empirical studies and experimentation. However, there is an unfortunate tradition of some researchers treating such communities merely as data sources for research instead of ``living organisms'' with their own values, goals, and norms. Research studies of Wikipedia often encounter resistance from Wikipedia editors because community values are sometimes violated by research activities. 

To avoid these problems, and to ensure community values are well attended to, the team worked with stakeholders to develop a research protocol\footnote{See https://meta.wikimedia.org/wiki/Research:WikiProject\_Recommendation} that was acceptable to the community. Essentially, we developed our research plan on the Wikipedia platform. In other words, we were developing and deploying our algorithms not just for, but \textit{with}, the Wikipedia community.

\subsubsection{Step 4: Deploy the algorithms and iteratively refine them.} 
We iteratively tested and refined our algorithms. Over a duration of six months we sent weekly batches of recommendations to our pilot participants, conducted short surveys to seek their views on these recommendations, and also interviewed project organizers and newcomers. We used their feedback to make significant changes to the algorithms. Example revisions included refining algorithm explanations and boosting the threshold of the matching algorithms. In particular, candidate editors will only be recommended by the rule-based matching algorithm if
they have made minimally five edits on project-related articles in the previous month.

\subsubsection{Step 5: Evaluate algorithms based on stakeholder acceptance, algorithm accuracy, and impacts of the algorithm on the community.} 
We conducted qualitative and quantitative studies to systematically examine the acceptance, accuracy, and impacts of the algorithms. The goal was to fully understand the influence of the algorithmic tool on the community and stakeholders, and to monitor any unintended consequences. 

Results from our initial evaluation included: 

\begin{enumerate}
    \item \textbf{Assessing acceptance among stakeholders}. We conducted interviews with community stakeholders, including newcomers and WikiProject organizers who used our recommendation systems. The feedback was positive. This gave us confidence that our system was acceptable to the stakeholders. For example, one organizer wrote: ``This puts some science behind recommendations, and will be a great supplement to the current processes.'' Editors who were invited to join projects also reacted positively. For example, an editor who was invited to join \textit{WikiProject Africa} wrote: ``Thank you for reaching out to me and thank you for informing me about the WikiProject Africa ... I appreciate it.''
    \item \textbf{Assessing accuracy}. Organizers rated the rule-based algorithm higher compared to the other three types of algorithms that were category-based, bonds-based and co-edit based. The average rating for the rule-based algorithm was 3.24 in a 5-point scale, significantly higher than the other three types of algorithms (\textit{t} = 3.51, \textit{p} <.001). The invitation rate (which is analogous to the click-through rate) for the rule-based algorithm was 47\%, which was also higher than category-based (16\%), bonds-based (22\%), and co-edit based (28\%). There was no significant difference among the other three algorithms. We also found that inexperienced editors and experienced editors were equally invited by the project organizers through the recommendation system. 
    \item \textbf{Assessing impacts}. Our evaluation of the impacts on the community sought to understand what happened to newcomers if WikiProject organizers invited them to the projects. Initial results suggested that: (1) Only experienced editors who received invitations from project organizers had a significant increase in their within-project activities over the baseline group composed of equally competent editors who did not receive invitations; and (2) the increase in the invited experienced editors' contributions did not come at the cost of fewer edits of Wikipedia articles beyond the joined WikiProject \cite[see][]{Zhu2018Value}.
\end{enumerate}

\subsection{Discussion and Implications}
\subsubsection{Challenges and Future Directions}

While this WikiProjects case demonstrated the promise of Value Sensitive Design and the five-stage holistic approach, it also helped reveal several challenges. First, it is challenging to explain the algorithms in ways that enable stakeholders to assess them and to provide sensible feedback to improve them. This constitutes a barrier to full stakeholder engagement in the design process. One future direction is to explore the effectiveness and trade-offs of different strategies and interfaces to help non-expert stakeholders understand algorithmic decision-making.  

Second, there is a lack of holistic understanding of the mapping between a wide range of human values and a variety of algorithmic design options. One promising direction for future work is to design ways to translate human values into different algorithmic choices, covering areas such as data pre-processing, model regularization during the learning process, model post-processing, and model selection.

\subsubsection{Implications for Learning Analytics System Design}

This study has a number of implications for learning analytics design. First, the designed algorithm products can be transferred to other learning contexts. For instance, given the increasing popularity of virtual collaboration in MOOCs \cite{Wen2015-oh}, we could design similar recommendation algorithms to support the formation of productive virtual learning groups in MOOCs. We have not seen work in this area attending to learner values yet, and the Value Sensitive Design methodology applied to the Wikipedia context could contribute to settings like MOOCs.

Second, the five-stage design process is directly applicable to a learning analytics design project. Even though the study was primarily focused on algorithms, the Value Sensitive Design process introduced in this study can be applied to the design of a learning analytics system, of which algorithm is an important component \cite{Buckingham_Shum2017-un}. While we could embrace the process when choosing or developing network visualization layout algorithms for a social network analytics tool, we can also apply the process to other parts of the analytics system. For example, as demonstrated in this case study, the consideration of human values is critical for the process of communicating analytics results to stakeholders. Other aspects of the system, such as the technological infrastructure used to store and transit learning data can also benefit from the Value Sensitive Design approach.

\section{General Discussion and Conclusions}

To support ethical considerations and system integrity in learning analytics, this paper introduces the application of Value Sensitive Design in learning analytics design processes. The first study was a conceptual investigation of an existing learning analytics tool. Applying Value Sensitive Design methods---specifically stakeholder analysis and value analysis---we uncovered a number of values (e.g. student autonomy, self-image) and value tensions with the studied tool. Based on these analyses, we proposed design trade-offs that are subject to further empirical, technical, and conceptual investigations. 

The second study went a step further. It proposed a holistic process of conducting Value Sensitive Design and demonstrated its application to building recommendation systems for WikiProjects---an informal, mass collaboration scenario. The design process includes literature analysis, empirical investigations, prototype development, community engagement, iterative testing and refinement, and continuous evaluation. Even though this study was focused on algorithms, the process can be readily applied to the design of a whole learning analytics system.

Overall, this paper is motivated by a need of practical means to support ethical considerations and human values in learning analytics system. Learning analytics operates in the ``middle space'' between learning and analytics and needs to consider a wide range of factors including learning theory, data integrity, human--computer interaction, and sociocultural factors \cite{Suthers2013-qm}. As a result, a wide range of values, which go beyond ethical concerns and are often competing with each other, need to be considered. 

Value Sensitive Design is a promising approach to promoting ethical practice in learning analytics. Preliminary work in this area has already been reported by colleagues \cite{Verhagen2016-sf}, and more systematic push for the application of Value Sensitive Design in learning analytics is needed. Two studies reported in this paper mitigate the gap. They demonstrate two ways of adopting Value Sensitive Design in learning analytics projects: one was a conceptual investigation (a ``lite'' version), whereas the other was a detailed design process comprising multiple Value Sensitive Design methods (a ``holistic'' version). It is hoped both versions of using Value Sensitive Design can strengthen future design of learning analytics systems. 

To improve learning analytics systems (esp. its ethical dimension), algorithmic transparency and accountability are not enough; rather, we should consider the system's integrity holistically \cite{Buckingham_Shum2017-un}. This paper is only an initial step towards this direction. Much work remains to be done to infuse Value Sensitive Design more broadly in the field of learning analytics. Interestingly, several projects have emerged to advance this work. Besides the \textit{Privacy by Design} approach mentioned above \cite{Verhagen2016-sf}, the Connected Intelligence Centre at the University of Technology Sydney is working on an \textit{Ethical Design Critique} (EDC) approach to involve stakeholders to consider ethical and risk factors involved in new learning analytics tools. Even though this approach does not directly apply Value Sensitive Design, it considers competing values identified by different stakeholders and aims to reach resolutions. One future direction can be purposefully exploring synergies among algorithm auditing, Privacy by Design, Ethical Design Critique, and Value Sensitive Design to comprehensively consider human values in learning analytics.

\bibliographystyle{ACM-Reference-Format}
\bibliography{lak19-bibliography}

\end{document}